\documentclass{article}
\usepackage{amssymb}

\input{tcilatex}
\begin{document}

\title{Theoretical Particle Limiting Velocity From The Bicubic Equation:
Neutrino Example}
\author{Josip \v{S}oln \\
JZS Phys-Tech, Vienna, Virginia 22182\\
E mail: soln.phystech@cox.net}
\date{ January 2014}
\maketitle

\begin{abstract}
There has been a lot of interest in measuring the velocities of massive
elementary particles, particularly the neutrinos. Some neutrino experiments
at first observed superluminal neutrinos, thus violating the velocity of
light $c$ as a limiting velocity. But, after eliminating some mistakes, such
as, for the OPERA\ experiments plugging the cable correctly and calibrating
the clock correctly, the measured neutrino velocity complied with $c$.
Pursuing the theoretical side of particle limiting velocities, here directly
from the special relativistic kinematics, in which all physical quantities
are in the overall mathematical consistency with each other, one treats
formally the velocity of light $c$ as yet to be deduced particle limiting
velocity, and derives the bicubic equation for the particle limiting
velocity in the arbitrary reference frame. The Lorentz invariance (LI) of
the energy-momentum dispersion relation assumes the velocity of light $c$ to
be universal limiting velocity of any particle. This expects the physical
solutions of the bicubic equation to be constrained in a sense that any
physical limiting velocity solution should equal numerically to $c$, \
preferably exactly, or at least, in a extremely good approximation. A rather
large numerical departure from $c$ means solution which, if it is physical,
would indicate the significant degree of Lorentz violation (LV). However,
this LV could be false if experimentally particle parameters were read
wrongly yielding different from $c$ solution for the physical limiting
velocity. Still, one may allow possible anticipation of finding some LV in
neutrino physics. The three solutions for the squares of limiting
velocities, denoted as $c_{1}^{2},c_{2}^{2}$ and $c_{3}^{2}$, depend on the
particle $m$, $E$ and $v$ (mass, energy and ordinary velocity) through
inverse sinusoidal functions. As $c_{1}^{2}$, $c_{3}^{2}>0$ and $c_{2}^{2}$ $%
<0,$ only $c_{1}^{2}$ and $c_{3}^{2}$ have chances to be physical while $%
c_{2}^{2}$ is unphysical. Furthermore, with the inverse sinusoidal functions
principal values dependences, $c_{1}^{2}$ and $c_{3}^{2}$ being positive are
complementary limiting velocity squares, at least one of them physical and
presumed luminal, while $c_{2}^{2}$ being negative is definitely unphysical.
However, $c_{2}^{2}$ can become $c_{1}^{2}$ and $c_{3}^{2}$ when transformed
from the principal values region into the multiple values region of the
inverse sinusoidal functions. With these solutions one can treat physical
limiting velocities, for any particle, electron, neutrino, photon, etc. The
OPERA $17GeV$ muon neutrino velocity experiments are discussed through the
limiting velocity $c_{3}$ because the calculated neutrino $m_{\nu }c^{2}$ of 
$0.076$ $eV$ , being negligible, makes $c_{1}$ unphysical. Furthermore,
because in OPERA experiments, $m_{\nu }c^{2}<<E_{\nu }$, one finds out that $%
c_{3}=c(1+\Delta )\simeq c$ \ because $\Delta $ is negligible (it varies
from $O(-10^{-6})$ to $O(10^{-6})$ ). This implies basically the LI of the
neutrino energy-momentum dispersion relation.
\end{abstract}

\textbf{1. Introduction}

\bigskip

\bigskip

There have been the whole series of neutrino velocity experiments such as
the OPERA collaborations with the detector in the CNGS beam [1] (versions 1,
2 and 3), the OPERA detector the CNGS beam using the 2012 dedicated data [2]
(versions 1 and 2) as well as the ICARUS detector in the CNGS beam [3]
(versions 1 and 2).

These neutrino velocity experiments are not simple to carry out and a lot of
them had a variety mistakes . For instance in [1] (version 1) a cable was
incorrectly plugged and there was a miscallibrated atomic clock. Both
mistakes were found and the OPERA\ collaborators made proper corrections and
after a precise measurment of the neutrino velocity in agreement with $c$,
the velocity of light, published the result in [1] (version 3). OPERA
collaborators also published the results from a measurement of a special
bunched neutrino beam [2] (version 2) giving the precision measurement of
the muon neutrino and muon anti-neutrino velocities, in good agreement with $%
c$, the velocity of light. Other, so called Gran Sasso laboratory repeated
the measurements and obtained $c$, the velocity of light, for the neutrino
velocity [3] (version 2).\ Now, the masses of flavor neutrinos, whose
velocities one can measure, are not yet known precisely but are calculated
as exactly as possible from the provided masses of the mass state neutrinos.
Nevertheless, the accepted notion from the special relativity, also in cases
like these, expects the neutrino velocity not to exceed the velocity of
light $c$, considered in the special relativity as the universal limiting
velocity.

In Section 2,from relativistic kinematics one formulates the sixth order
bicubic equation for the square of the limiting velocity and, as exposed in
[4], can be solved as a cubic equation for $c^{2}$. The bicubic equation
yields three solutions $c_{i}^{2},$ $i=1,2,3$, depending on $m,v,$ and $E$
(particle mass, velocity and energy). One expects that at least one solution
is physical and luminal and as such supports the LI; that is when evaluated
to be numerically, either exactly or practically exactly equal to $c$, so
that its substitution in place of $c$, will not change at all the
energy-momentum dispersion relation or it will change it insignificantly.
The three limiting velocity solutions depend on the inverse sinusoidal
function principal values in such a way that the complementary $c_{1}$ and $%
c_{3}$ are real and, at least one of them, physical while $c_{2}$ is
imaginary and as such unphysical. However, for the specific multiple values
of the inverse sinusoidal function, $c_{2}$ can become $c_{1}$ and $c_{3}$.
The important thing is the fact that $c_{1}$ and $c_{3}$ are complementary
limiting velocities since they together can cover all the allowed particle
parameters, $m,v,$ and $E$ , while each of them is limited to particular
values. Besides the exact solutions, Section 2 \ contains also the
perturbative solutions for $c_{i}^{2},$ $i=1,2,3$, basically in terms of $%
(mv^{2}/E)$ . These perturbative solutions are often very convenient for
determining as to which of the limiting velocities is physical, $c_{1}$ or $%
c_{3}$ , either luminal $(=,\simeq c)$ or not $(\neq c)$ that is, applicable
for the particle in question.

In Section 3, after deducing the three flavor neutrino masses, one finds out
that the physical parameter structure of the OPERA [2] muon neutrino
velocity experiment is such that the Taylor series expansion, in terms of $%
(mv^{2}/E)$, strongly suggests $c_{3}$ as the luminal solution. Along the
same lines, one notices that $c_{1}$ is unphysical\ \ in the OPERA [2]
experiments.The same is true for other neutrino velocity experiments [1] and
[3]. Furthermore, because the muon neutrino mass being negligible, one finds
out that in fact $c_{3}\simeq c$. To verify this perturbative result, one
performs the calculation also with exact non-perturbative expression for $%
c_{3}$. The result is the same as from the pertubative calculation.

Conclusion and final remarks are given in Section 4. Here also the
comparisons with other approaches from the literature for discussing the
Lorentz invariance and Lorentz violation, either through changes in the
Dirac equation or by explicitly changing the relativistic kinematics, are
given.

\bigskip

\textbf{2. Particle limiting velocities from the velocity bicubic equation}

\bigskip

The velocity of light, $c$, in the special relativity particle kinematics is
considered the universal relativistically invariant limiting velocity. Here,
with the desire of having $c$ on an equal basis with other physical
parameters, one treats it as yet to be analythically formulated limiting
velocity and starts with the same kinematics.

\begin{equation}
\overrightarrow{p}=\frac{E\text{ }\overrightarrow{v}}{c^{2}},E^{2}=\frac{%
m^{2}c^{4}}{1-\frac{v^{2}}{c^{2}}}  \tag{1,2}
\end{equation}%
which defines the particle momentum $\overrightarrow{p}$, and energy $E,$
through its mass $m$ and velocity $\overrightarrow{v}$. Momentum and energy
from (1) and (2) are related through the mass shell condition,

\begin{equation}
\overrightarrow{p}^{2}c^{2}-E^{2}=-m^{2}c^{4}  \tag{3.1}
\end{equation}%
whose change, if any, caused by replacing $c$ with limiting velocity
solutions from the bicubic equation (to be discussed), could indicate either
LI or LV, providing that particle parameters, $m,v,$ and $E$ are known.

As in the neutrino velocity experiments [1,2,3], one has the known energy
and the velocity of fixed direction, it is convenient to continue with just
relation (2) by transforming it into the bicubic equation for the particle
limiting velocity $c:$

\begin{equation}
m^{2}c^{6}=E^{2}c^{2}-E^{2}v^{2}  \tag{3.2}
\end{equation}%
Next, one rewrites it in the mathematically more familiar forms with
solutions characterized \ by the discriminant $\ $satisfying $D<0$ ,

\begin{eqnarray}
\left[ \left( \frac{c}{v}\right) ^{2}\right] ^{3}-\left( \frac{E}{mv^{2}}%
\right) ^{2}\left( \frac{c}{v}\right) ^{2}+\left( \frac{E}{mv^{2}}\right)
^{2} &=&0,\text{ \ }q=-p=\left( \frac{E}{mv^{2}}\right) ^{2},\text{ } 
\nonumber \\
D &=&\left( \frac{q}{2}\right) ^{2}+\left( \frac{p}{3}\right) ^{3}  \nonumber
\\
&=&\frac{1}{4}\left( \frac{E}{mv^{2}}\right) ^{4}\left[ 1-\frac{4}{27}\left( 
\frac{E}{mv^{2}}\right) ^{2}\right] <0,  \nonumber \\
&&  \TCItag{3.3}
\end{eqnarray}

\begin{equation}
z=\frac{3\sqrt{3}mv^{2}}{2E};\text{ }D<0:\text{ }-1<z<1  \tag{4}
\end{equation}%
According to [4], the solutions for (3.2,3) plus (4) can be written as

\begin{eqnarray}
c_{1}^{2} &=&2v^{2}\sqrt{\frac{\left\vert p\right\vert }{3}}\cos \left( 
\frac{\theta }{3}+\frac{\pi }{6}\right) ,\text{ }  \nonumber \\
c_{2}^{2} &=&-2v^{2}\sqrt{\frac{\left\vert p\right\vert }{3}}\cos \left( 
\frac{\theta }{3}-\frac{\pi }{6}\right) ,\text{ }  \nonumber \\
c_{3}^{2} &=&-2v^{2}\sqrt{\frac{\left\vert p\right\vert }{3}}\cos \left( 
\frac{\theta }{3}+\frac{\pi }{2}\right)  \nonumber \\
\cos \left( \theta +\frac{\pi }{2}\right) &=&-\frac{q}{2\left( \frac{%
\left\vert p\right\vert }{3}\right) ^{\frac{3}{2}}}=\frac{-3\sqrt{3}mv^{2}}{%
2E},  \nonumber \\
\text{ }\theta &=&-\frac{\pi }{2}+\cos ^{-1}\left( \frac{-3\sqrt{3}mv^{2}}{2E%
}\right) =\sin ^{-1}\left( \frac{3\sqrt{3}mv^{2}}{2E}\right)  \TCItag{5}
\end{eqnarray}%
However, in order to see more of the physics, the exact solutions from (5),
with the help from relations (3.3) and (4), are rewritten in such a way as
to exhibit \ more explicitly the $m$, $v$, and $E$ parameters in them:

\begin{eqnarray}
c_{1}^{2} &=&\frac{2E}{\sqrt{3}m}\sin \left[ \frac{\pi }{3}-\frac{1}{3}\sin
^{-1}\left( \frac{3\sqrt{3}mv^{2}}{2E}\right) \right] >0,  \TCItag{6.1} \\
c_{2}^{2} &=&-\frac{2E}{\sqrt{3}m}\cos \left[ \frac{1}{3}\sin ^{-1}\left( 
\frac{3\sqrt{3}mv^{2}}{2E}\right) -\frac{\pi }{6}\right] <0,  \TCItag{6.2} \\
c_{3}^{2} &=&\frac{2E}{\sqrt{3}m}\sin \left[ \frac{1}{3}\sin ^{-1}\left( 
\frac{3\sqrt{3}mv^{2}}{2E}\right) \right] \text{ }>0  \TCItag{6.3}
\end{eqnarray}

Noting that with the variable $z$ from relation (4) \ the inverse sinus
function, $\sin ^{-1}\left( z\right) $,$\ $in (6.1,2,3) refer to the
principal values that lie in the $\left( -\pi /2\ \ to\ \ \pi /2\right) $
range where either of the positive $c_{1}^{2}$ and $c_{3}^{2}$ \ can be
physical while the negative $c_{2}^{2}$ is definitely unphysical. However, $%
c_{2}^{2}$ can become physical in the multiple values ranges. Denote $%
c_{i}^{2},i=1,2,3$ dependence on $z$ as $c_{i}^{2}\left[ \sin ^{-1}\left(
z\right) \right] ,i=1,2,3$ . Then assume that alternately in $c_{2}^{2}$ the
range of $\sin ^{-1}\left( z\right) $ is changed from $\left( -\pi /2\ to\ \
\pi /2\right) $ to $(3\pi /2$ $to$ $5\pi /2)$ and to $(\left( -5\pi
/2\right) to$ $\left( -3\pi /2\right) ).$This is simply achieved by
replacing $c_{2}^{2}\left[ \sin ^{-1}\left( z\right) \right] $ in (6.2)
alternately with $c_{2}^{2}\left[ \sin ^{-1}\left( z\right) +2\pi \right] $
and $c_{2}^{2}\left[ \sin ^{-1}\left( z\right) -2\pi \right] $. Then the
simple evaluations, with the help from (6.1,2,3), shows that

\begin{equation}
c_{2}^{2}\left[ \sin ^{-1}\left( z\right) +2\pi \right] =c_{3}^{2}\left[
\sin ^{-1}\left( z\right) \right] ;\text{ }c_{2}^{2}\left[ \sin ^{-1}\left(
z\right) -2\pi \right] =c_{1}^{2}\left[ \sin ^{-1}\left( z\right) \right] 
\tag{6.4,5)}
\end{equation}%
In (6.4,5) the respective multivalue ranges in $c_{2}^{2}$ are $\left( 3\pi
/2\ \ to\ \ 5\pi /2\right) $ and $\left( -5\pi /2\ \ to\ -\ 3\pi /2\right) $
while in $c_{3}^{2}$and $c_{1}^{2}$ the range is $\left( -\pi /2\ \ to\ \
\pi /2\right) $. As one sees , here the complementary limiting velocities $%
c_{1}$ and $c_{3}$ are the only ones of the physical significances.
Importance of (6.4,5) is in the fact that sometimes one has to change the
"coordinates" in order to find out the same or perhaps even the new physics.
These different ranges, principle values and multiple values do not change
the fact that according to (4) $|z|<1$. The imaginary $c_{2}$ moves the
"imaginary" physics to the "real" physics with $\sin ^{-1}\left( z\right) $,
being changed to $\sin ^{-1}\left( z\right) \pm 2\pi $ now with the multiple
value ranges.

Next, it is illustrative to perform the Taylor series expansions of $%
c_{i}^{2},i=1,2,3$ (6.1,2,3) . Except for the few first terms, they are done
basically in terms of $(mv^{2}/E)$ with inequalities between $v,E$ and $m$
as indicated in each of the series. Extra $v^{2}$ factor in each term makes
the whole expression to have dimension of $v^{2}$ .

\begin{eqnarray}
c_{1}^{2} &=&\frac{E}{m}-\frac{v^{2}}{2}-\frac{3mv^{4}}{8E}-\frac{m^{2}v^{6}%
}{2E^{2}}+O\left[ v^{2}\left( \frac{mv^{2}}{E}\right) ^{3}\right] >0,v^{2}%
\text{ }<\frac{2E}{3\sqrt{3}m},  \TCItag{7.1} \\
c_{2}^{2} &=&-\frac{E}{m}-\frac{v^{2}}{2}+\frac{3mv^{4}}{8E}-\frac{m^{2}v^{6}%
}{2E^{2}}+O\left[ v^{2}\left( \frac{mv^{2}}{E}\right) ^{3}\right] <0,v^{2}%
\text{ }<\frac{2E}{3\sqrt{3}m},  \TCItag{7.2} \\
c_{3}^{2} &=&v^{2}+\frac{m^{2}v^{6}}{E^{2}}+\frac{69}{32}\frac{m^{4}v^{10}}{%
E^{4}}+O\left[ v^{2}\left( \frac{mv^{2}}{E}\right) ^{6}\right] >0,m<\frac{2E%
}{3\sqrt{3}v^{2}}  \TCItag{7.3}
\end{eqnarray}

Already from exact solutions (6.1,2,3) as well as now from the Taylor
series, one sees that all three limiting velocities are different from each
other in the principle values region. For instance, $c_{1}^{2}$ and $%
c_{2}^{2}$ diverge for $m=0$ but are finite for $v=0$. So $c_{1}^{2}$ and
the unphysical $c_{2}^{2}$ need to have $m\neq 0$. The different behaviors
of $c_{1}^{2}$ and $c_{3}^{2}$ for either $m\rightarrow 0$ or $v\rightarrow
0 $ emphasizes their complementarity. This small excursion, suggests
defining the physical $c_{1}^{2}$ and $c_{3}^{2}$ in the principle values
region satisfying

\begin{equation}
Phyhsical:c_{1}^{2},\text{ }c_{3}^{2}\neq 0,\infty  \tag{8.1}
\end{equation}%
Here, consistent with (8.1), is the summary of important situations that can
occur for $c_{1}^{2}$ and $c_{3}^{2}$%
\begin{eqnarray}
v &=&0,m\neq 0:\frac{E_{0}}{m}=c_{1}^{2}=c^{2};\text{ }c_{3}^{2}=0\text{ }%
(unphysical),  \TCItag{8.2} \\
m &=&0,E\text{ }finite;c_{3}=v=c\text{ }(photon);\text{ }c_{1}=\infty \text{ 
}(unphysical),  \TCItag{8.3} \\
m &\neq &0,E\text{ }\rightarrow \infty ;c_{3}\rightarrow v;\text{ }%
c_{1}\rightarrow \infty \text{ }(unphysical).  \TCItag{8.4}
\end{eqnarray}%
What examples (8.2,3,4) show clearly is that in these particular situations
the physically acceptable limiting velocity is either given by $c_{1}$ or $%
c_{3}$ which further indicates to their complementarity. The relation (8.2)
is to be understood as a definition of $E(v)$ at $v=0$ where $%
c_{1}^{2}=c^{2} $. As $c_{1}$ and $c_{3}$ are the limiting velocity
solutions of the bicubic equation (3.3), it is appropriate to see what
effect will be caused if one sets either $c_{1}$ or $c_{3}$ in place of $c$
in the energy momentum relation (3.1). These substitutions leave the energy
momentum relation (3.1) either LI or, to a degree, LV under the Lorentz
transformations with the following respective general possible values for $%
c_{1}$ or $c_{3}$:

\begin{equation}
LI:c_{1}=c\text{ }or\text{ }c_{3}=c;\text{ }LV:c_{1}\neq c\text{ }or\text{ }%
c_{3}\neq c  \tag{8.5}
\end{equation}%
Here it is assumed that either $c_{1}$ or $c_{3}$ is LI but not both of them
at the same time. Also, the degree of the LV would depend on how strongly $%
c_{1}\neq c$ $or$ $c_{3}\neq c$.

Despite their complementarity, It is necessary to investigate whether it can
happen that for a given particle one can have $c_{1}=c_{3}$ ? Imposing this
equality, from (6.1) and (6.2), with $z$ as defined in (4), one arrives at
the following sequence of equations,

\begin{eqnarray}
c_{1}^{2} &=&c_{3}^{2}:\sin \left[ \frac{\pi }{3}-\frac{1}{3}\sin
^{-1}\left( z\right) \right] =\sin \left[ \frac{1}{3}\sin ^{-1}\left(
z\right) \right] ,  \TCItag{9.1} \\
\sin ^{-1}\left( z\right) &=&\frac{\pi }{2}:\text{ }z=1  \TCItag{9.2}
\end{eqnarray}

Since relation (9.2) is in contradiction to the relation (4), which excludes 
$z=1$, one concludes that $c_{1\text{ }}$and $c_{3}$, while complementary,
cannot be equal to each other for the same particle,$c_{1}\neq c_{3}$ .
Hence, if for instance $c_{3\text{ }}=c$ then $c_{1}\neq c$ and so on.\ \ \
\ \ \ \ \ \ \ \ \ \ \ \ \ \ \ \ \ \ \ \ \ \ 

\bigskip

\textbf{3. Limiting velocity of the neutrino}

\bigskip

Here one is specifically interested in applying the formalism of obtaining
the limiting velocity for the muon neutrino, $\nu _{\mu }$, with the
physical parameters from the OPERA experiment [2]. From the perturbative
expressions (7) , the indication is that $c_{1}$ and $c_{3}$ are
respectively, the unphysical and physical limiting velocities, in this case.
To see whether the physical $c_{3\text{ }}$is also luminal, that is, leading
to $c$ and LI, one first expresses $c_{3}$ perturbatively from (7.3) and
then, for verification purposes, also exactly from (6.3).

The formalism in relations (6) and (7) demand, in addition to $E$ and $v$
also the value of the mass, here denoted for the muon neutrino as $m_{\nu }$%
. As in the reference [2] the value of $m_{\nu }$ is not given, one has to
first find which value is presently favored, although in OPERA experiment
[2] with the neutrino energy of $\ E_{\nu }\left( \mu \right) =17GeV$, the
calculated neutrino mass even if exact, will be very likely negligible as
compared to the energy. Now,there are three flavor neutrinos, denoted as $%
\nu _{e}$, $\nu _{\mu }$ and $\nu _{\tau }$, the electron, muon and tau
neutrino.Their masses $m_{\nu }(e)$, $m_{\nu }(\mu )$ and $m_{\nu }(\tau )$
are derived from the masses of the independent mass-state three neutrinos
with masses $m_{1}$,$m_{2\text{, }}$and $m_{3}$. In the discussion of the $%
\mu -\tau $ symmetry, these masses have been given by S. Gupta et al. [5] as,

\begin{equation}
m_{1}c^{2}=0.067\text{ }eV,\text{ }m_{2}c^{2}=0.068\text{ }eV,\text{ }%
m_{3}c^{2}=0.084\text{ }eV\text{,}  \tag{10.1}
\end{equation}%
The flavor neutrino masses are defined in [5] with the help of the Harrison
et al. neutrino mixing matrix [6], $U_{\alpha ,i},$ $\alpha =e,\mu .\tau $ $%
;i=1,2,3$, connecting the flavor neutrino states to the mass-state neutrino
states (see,also [7] ]). Hence, using $U_{\alpha ,i}$ as in references [6]
and [7], according to Gupta et al. [5], the flavor neutrino masses are
defined as

\begin{eqnarray}
\alpha &=&e,\upsilon ,\tau ;\text{ }i=1,2,3:\text{ }m_{\nu }\left( \alpha
\right) =\left[ \tsum\nolimits_{i}\left\vert U_{\alpha ,i}\right\vert
^{2}m_{i}^{2}\right] ^{\left( \frac{1}{2}\right) };  \nonumber \\
\text{ }\left( U_{\alpha ,i}\right) &=&\left( 
\begin{tabular}{lll}
$\sqrt{\frac{2}{3}}$ & $\sqrt{\frac{1}{3}}$ & $0$ \\ 
$-\sqrt{\frac{1}{6}}$ & $\sqrt{\frac{1}{3}}$ & $-\sqrt{\frac{1}{2}}$ \\ 
$-\sqrt{\frac{1}{6}}$ & $\sqrt{\frac{1}{3}}$ & $\sqrt{\frac{1}{2}}$%
\end{tabular}%
\right)  \TCItag{10.2}
\end{eqnarray}%
yielding

\begin{equation}
m_{\nu }\left( e\right) c^{2}=0.067\text{ }eV,\text{ }m_{\nu }\left( \mu
\right) c^{2}=0.076\text{ }eV,\text{ }m_{\nu }\left( \tau \right) c^{2}=0.076%
\text{ }eV  \tag{11}
\end{equation}%
\ With these values and from [2] the collection of data for the OPERA muon
neutrino velocity experiment is

\begin{eqnarray}
E_{\nu }\left( \mu \right) &=&17GeV,\text{ }m_{\nu }\left( \mu \right)
c^{2}=0.076\text{ }eV,\text{ }  \nonumber \\
v_{\nu }\left( \mu \right) &=&c(1+\Delta )  \nonumber \\
-1.8\times 10^{-6} &\leq &\Delta \leq 2.3\times 10^{-6}  \TCItag{12}
\end{eqnarray}

Because one has that $\ m_{\nu }\left( \mu \right) <\left( 2E_{\nu }\left(
\mu \right) /3\sqrt{3}v_{\nu }^{2}\left( \mu \right) \right) $ for any $%
v_{\nu }\left( \mu \right) $ from (12), one easily deduces, according to
(7.1), that approximate numerical value of $c_{1}$ is $\ c_{1}$ $\simeq
4.73\times 10^{5}c$ .Such a large value makes $c_{1}$ unphysical. Expecting
that $c_{3}$ is physical, one calculates it with more precision first
perturbatively from (7.3),

\begin{equation}
c_{3}^{2}=v_{\nu }^{2}\left( \mu \right) \left[ 1+\left( \frac{m_{\nu
}\left( \mu \right) c^{2}}{E_{\nu }\left( \mu \right) }\right) ^{2}\left( 
\frac{v_{\nu }\left( \mu \right) }{c}\right) ^{4}+O\left( \left( \frac{%
m_{\nu }\left( \mu \right) c^{2}}{E_{\nu }\left( \mu \right) }\right)
^{4}\left( \frac{v_{\nu }\left( \mu \right) }{c}\right) ^{8}\right) \right] 
\tag{13}
\end{equation}%
Furthermore with $\left( m_{\nu }\left( \mu \right) c^{2}/E_{\nu }\left( \mu
\right) \right) \simeq 4.5\times 10^{-12}$ \ and $|\Delta |<<1$, one obtains
for $c_{3}$ the perturbative solution in the form,

\begin{equation}
\frac{c_{3}}{c}\simeq \left( 1+\Delta \right) \left( 1+\frac{1}{2}\left( 
\frac{m_{\nu }\left( \mu \right) c^{2}}{E_{\nu }\left( \mu \right) }\right)
^{2}(1+\Delta )^{4}\right) =\left( 1+\Delta \right)  \tag{14.1}
\end{equation}

Exact expression for $c_{3}$ from (6.3) with the OPERA physical parameters
is as follows

\begin{eqnarray}
\left( \frac{c_{3}}{c}\right) ^{2} &=&\frac{2E_{\nu }\left( \mu \right) }{%
\sqrt{3}m_{\nu }\left( \mu \right) c^{2}}\sin \left[ \frac{1}{3}\sin
^{-1}\left( \frac{3\sqrt{3}m_{\nu }\left( \mu \right) c^{2}\left( 1+\Delta
\right) ^{2}}{2E_{\nu }\left( \mu \right) }\right) \right]  \nonumber \\
&=&(1+\Delta )^{2}  \TCItag{14.2}
\end{eqnarray}%
In deriving (14.2), one simply takes into account that $\left( m_{\nu
}\left( \mu \right) c^{2}/E_{\nu }\left( \mu \right) \right) <<1$ and that
in $(1+\Delta )$ ,\ $|\Delta |<<1$ so that only first terms in Taylor
expansions of sin and sin$^{-1}$ functions need to be retained. Hence , both
solutions, being equal and basically luminal, yield LI with $c$ as the
solution for $c_{3}$:$\ \ \ \ \ \ \ \ \ \ \ \ \ \ \ \ \ \ \ \ \ \ \ \ \ \ \
\ \ \ \ \ \ \ \ \ \ \ \ \ \ \ \ \ \ \ \ \ \ \ \ \ \ \ \ \ \ \ \ \ \ \ \ \ \
\ \ \ \ \ \ \ \ \ \ \ \ \ \ \ \ \ \ \ \ \ \ \ \ \ \ \ \ \ \ \ \ \ \ \ $%
\begin{equation}
\ \ \ \ \ \ \ (1-1.8\times 10^{-6})c\leq \ c_{3}\leq (1+2.3\times 10^{-6})c\
:\text{ }c_{3}\simeq c  \tag{14.3}
\end{equation}%
The result in (14.3) is what Einstein envisioned long time ago.

\ \ \ \ \ \ What one notices here is the fact that for OPERA experiments
[2], through a particular collection of neutrino physical parameters, such
as mass, ordinary velocity and energy, the bicubic equation yields the
luminal limiting velocity solution, that is, with the velocity of light $c$
and with the LI.

Now, on one example one can show how the luminal limiting velocity solution
with the LI, can become the superluminal solution with the LV. Simply, in
(14.1) and (14.2) replace the negligible $\Delta $ with a small but finite
and positive $\Delta $. In doing so, one basically obtains he LV from
reference [8],implied by the change in the particle special relativistic
velocity, written as $c/\left( p^{2}+m^{2}c^{2}\right) ^{\frac{1}{2}}+\Delta
c.$It is easily seen that in the situation where the mass is negligible, as
is in the OPERA experiments, this $\Delta $ should be the same as $\Delta $
in relations (14.1,2), and if negligible, as in relations (14), should allow
the LI rather than the LV under the Lorentz transformations. Although so far
no verifiable LV showed up in neutrino physics, on should, nevertheless,
keep an open mind also for such a possibility with subluminal or
superluminal anticipations. The LV formulations with superluminal particles
through the Dirac equation have been done, for example, in [9] and [10].

\bigskip

\textbf{4. Conclusion and final remarks}

\bigskip

Identifying the velocity of light $c$ in the relativistic kinematics as a
limiting velocity yet to be determined, one is lead naturally to the bicubic
equation for the limiting velocity. Of the three resulting solutions one, $%
c_{2}$, is imaginary while two other solutions, $c_{1}$ and $c_{3}$, are
real and complementary with different emphasis on particle parameter
dependences. Of course, if the particle parameters choose, say $c_{1}$ $=c$
,then as argued in (8.1) to (8.4) $c_{3\text{ }}$will be unphysical. The
remarkable point in determining the limiting velocity of any particle from
the bicubic equation is that the particle physical parameters will yield for
it most likely $c$, no matter where one measures its mass, energy and the
ordinary velocity.

It appears that what one needs are the velocity experiments done with rather
a masssive particle which allow full participation of the particle mass in
determining of its limiting velocity . A natural candidate for such a
limiting velocity determination is the electron whose mass is very well
known and the energy can be chosen so as not to render the mass negligible.

\bigskip

\textbf{Acknowledgment}

\bigskip

The author is grateful to Dr. M. Dracos for informing him that the neutrino
beam has 17 GeV average energy in the OPERA\ \ experiments. Descriptions and
explanations of neutrino velocity experiments by an anonymous physicist
knowledgeable with OPERA and other experiments is gratefully acknowledged.

\bigskip

\textbf{References}

\bigskip

\bigskip \lbrack 1] \ T. Adam et al., "Measurement of neutrino velocity with
OPERA detector in the CNGS beam."; arXiv:1109.4897.

[2] T. Adam et al., "Measurement of neutrino velocity with the OPERA
detector in the CNGS beam using the 2012 dedicated data"; arXiV:1212.1276

(17 Dec. 2012), published in JHEP 1210, 093 (2012).

\bigskip \lbrack 3] \ M. Antonello et al., "Precision measurements of the
neutrino velocity with the ICARUS detector in the CNGS beam.";
arXiv:1208.2629 (26 Sept., 2012). Phys. Lett. B\textbf{713, 17 (2012).}\ \ \
\ \ .

[4] \ R. S. Burrington, Handbook of Mathematical Tables and Formulas,
McGraw-Hill, 1973; \ H. J. Bartsch, Taschenbuch Matematischen Formeln
(Handbook of Mathematical Formulas),

\ \ \ \ \ Verlag \ Harri Deutsch, 1975.

[5] \ \ S. Gupta, A. S. Joshipura and K. M. Patel, "How Good is mu-tau
Symmetry After Results on Non-Zero Theta(13)?"; arXiv: 1301.7130 (30 Jan.,
2013).

[6] \ P. F. Harrison, D. H. Perkins and W. G. Scott, Phys. Lett. B\textbf{\
530, }167 (2002)\textbf{.}

[7] \ Josip \v{S}oln, Phys. Scr. \textbf{80,} 025101 (2009); arXiv:
0908.1763.

\bigskip \lbrack 8] \ G. Amelino-Camelia, "Phenomenology of Philosophy of
Science": OPERA data ; arXiv: 1206.3554 (2012).\ 

[9] \ S. R. Coleman and S. L. Glashow, Phys. Rev. D\textbf{59}, 116008
(1999).\ \ 

[10] \ L. Zhou, B. Q. Ma, Astropart. Phys. \textbf{41, }24-27 (2013); arXiv:
1109-6097 (2013).

\end{document}